\documentclass{elsart-1}

 \usepackage{graphicx}
\usepackage{color}
\begin{document}

\begin{frontmatter}



\title{Finsler geometry modelling of phase separation in multi-component membranes}


\author{Satoshi Usui and Hiroshi Koibuchi}

\address{Department of Mechanical and Systems Engineering \\
  National Institute of Technology, Ibaraki College \\
  Nakane 866, Hitachinaka, Ibaraki 312-8508, Japan }
\ead{koibuchi@mech.ibaraki-ct.ac.jp}

%\author{Andrey Shobukhov}

%\address{Faculty of Computational Mathematics and Cybernetics, Lomonosov Moscow State University, 119991, Moscow, Leninskiye Gory, MSU, 2-nd Educational Building, Russia}

\begin{abstract}
A Finsler geometric surface model is studied as a coarse-grained model
for membranes of three components, such as zwitterionic phospholipid (DOPC), lipid (DPPC) and an organic molecule (cholesterol). To understand the phase separation of liquid-ordered
(DPPC rich) $L_o$ and liquid-disordered (DOPC rich) $L_d$,  we
introduce a binary variable $\sigma (=\pm 1)$ into the triangulated
surface model. We numerically determine that two circular and stripe
domains appear on the surface. The dependence of the morphological
change on the area fraction of $L_o$ is consistent with existing
experimental results. This provides us with a clear understanding of
the origin of the line tension energy, which has been used to
understand these morphological changes in three-component
membranes. In addition to these two circular and stripe domains, a
raft-like domain and budding domain are also observed, and the several
corresponding phase diagrams are obtained. 

\end{abstract}

\end{frontmatter}

% main text
%\section{}
%\label{}
\maketitle
%----------------------------------------------------------
\section{Introduction}
%----------------------------------------------------------
Membranes of multiple components, such as 1,2-dioleoyl-sn-glycero-3-phosphocholine (DOPC), dipalmitoylphosphatidylcholine (DPPC) and cholesterol,
are receiving widespread attention because of their applications in
many fields of science and technology, and numerous studies on the
morphological changes have been conducted
\cite{Sarah-PRL2005,Yanagisawa-PRL2008,Yanagisawa-PRE2010,Lipowsky-SM2009,Lipowsky-PRL1993,Lipowsky-PRE1996}. In
these membranes, morphological changes are induced by a phase
separation. Indeed, the phase separation causes domain formation and
domain pattern transition between the liquid-ordered ($L_o$) and the
liquid-disordered ($L_d$) phases. This domain pattern transition
accompanies the morphological changes, such as the two circular
domains, the stripe domain, the raft domain and the so-called budding
domain \cite{Lipowsky-PRL1993,Lipowsky-PRE1996}. The multiplicity of
components, as in a glass transition \cite{GJug-PM2004}, is essential
for such a variety of morphologies. To date, these morphologies have
been studied on the basis of the line tension energy
\cite{Yanagisawa-PRE2010,Lipowsky-SM2009} in the context of the Helfrich-Polyakov (HP) model for membranes \cite{POLYAKOV-NPB1986,HELFRICH-1973}. The line tension energy is defined on the domain boundary and has an important role in the morphological changes \cite{Yanagisawa-PRE2010,Lipowsky-SM2009}.

However, the origin of the line tension energy is not well understood. In fact, it is unclear what type of internal structure is connected to the line tension energy until now.   The problem that should be asked is where the line tension energy originates.
Therefore, in this paper, we clarify and discuss the microscopic origin of the line tension energy. 

To understand the origin of the line tension energy, we introduce new
degrees of freedom $\sigma (=\pm 1)$ to represent the $L_o$ and
$L_d$ phases. The Ising model Hamiltonian, which we call aggregation
energy, for the variable $\sigma$ is included in the general HP model
Hamiltonian, where the "general" HP model refers to the HP model with
a nontrivial surface metric $g_{ab}(\not=\!\delta_{ab})$.  Note that
the general HP model can be discretized on triangulated surfaces and
becomes well defined only when it is treated in the context of Finsler
geometry
\cite{Koibuchi-Sekino-2014PhysicaA,Matsumoto-SKB1975,Bao-Chern-Shen-GTM200}. Moreover,
note that our strategy towards the multi-component membrane in this paper is a coarse-graining of the detailed information on the chemical structures of DOPC, DPPC and  cholesterol and on the interaction between them with the help of the variable $\sigma$ and HP surface model. In addition, from the viewpoint of modelling, it is very natural to extend the Hp model to the general HP model for explaining the morphological changes in multi-component membranes. Indeed, the HP model is considered as a straight forward extension of the linear chain model for polymers \cite{Doi-Edwards-1986}.

The remainder of this paper is organized as follows.  In Subsection
\ref{cont-model}, we introduce the continuous Hamiltonian, which is
identical to the Polyakov Hamiltonian \cite{POLYAKOV-NPB1986}. In
Subsection \ref{disc-model}, we introduce the two-component surface
model, which is defined by including the aggregation energy in the
Hamiltonian of the FG surface model. The aggregation energy is defined
by the variable $\sigma$, which is introduced to label the triangles
with $L_o$ and $L_d$. The Monte Carlo (MC) technique is briefly
discussed in Section \ref{MC-technique}, and the MC results are
presented in Section \ref{results}. Finally, we summarize the results
in Section \ref{conclusion}. In Appendix \ref{surface-model}, we
describe the technical details of the FG modelling.  In Subsection
\ref{discrete-model}, the discretization of the continuous model
introduced in Subsection  \ref{cont-model} is described, and  a
discrete model is obtained. From this discrete model, we obtain the
model for two-component membranes  by imposing a constraint on the
metric function. In \ref{surface-orientation}, we show that the models
constructed in \ref{discrete-model} are ill defined in the conventional modelling and  that the models become well defined only in the context of Finsler geometry modelling.

%------------------------------------------
\section{Two-component surface model}\label{two-comp-model}
%------------------------------------------
%------------------------------------------
\subsection{Continuous surface model}\label{cont-model}
%------------------------------------------
We begin with a continuous surface model, which is defined by the Polyakov Hamiltonian or the Gaussian energy $S_1$ for membranes and the bending energy $S_2$ with a  metric $g(x)$, where $x\!=\!(x_1,x_2)$ is the local coordinate of the two-dimensional parameter space $M$ \cite{FDAVID-SMMS2004}. Both of the energies are defined by the surface position ${\bf r} (\in\Re^3)$ such that
\begin{eqnarray} 
\label{cont_S1S2}
&&S_1=\int \sqrt{g}d^2x g^{ab} \frac{\partial {\bf r}}{\partial x_a}\cdot \frac{\partial {\bf r}}{\partial x_b}, \nonumber \\ 
&&S_2=\frac{1}{2}\int \sqrt{g}d^2x  g^{ab} \frac{\partial {\bf n}}{\partial x_a} \cdot\frac{\partial {\bf n}}{\partial x_b}, 
\end{eqnarray} 
where $g$ is the determinant of the $2\!\times\!2$ matrix  $g_{ab}$ of
the metric function and $g^{ab}$ is its inverse
\cite{FDAVID-SMMS2004}. The symbol ${\bf n}$ denotes a unit normal
vector of the surface. Both $S_1$ and $S_2$ are conformally
invariant. The conformal invariance is a property in which a scale
change $g_{ab}(x)\!\to\! f(x)g_{ab}(x)$ is not reflected in both $S_1$
and $S_2$ for any positive function $f$.  Two metrics $g$ and
$g^\prime$ are called "conformally equivalent" if a function $f(x)$
exists such that  $g_{ab}^\prime\!=\!f(x)g_{ab}$ \cite{FDAVID-SMMS2004}.

For the case where $g_{ab}(x)$ is given by the Euclidean metric $g_{ab}\!=\!\delta_{ab}$ (or the induced metric $g_{ab}\!=\!\partial_a {\bf r}\cdot\partial_b {\bf r}$),  the surface shape ${\bf r}$ in $\Re^3$ is treated from the perspective of statistical mechanics. These are the HP model \cite{POLYAKOV-NPB1986,HELFRICH-1973} corresponding to polymerized membranes, and the HP model and the Landau-Ginzburg model \cite{PKN-PRL1988} have been thoroughly investigated \cite{KANTOR-NELSON-PRA1987,P-L-1985PRL,DavidGuitter-1988EPL,NELSON-SMMS2004,Bowick-PREP2001,WIESE-PTCP19-2000,GOMPPER-KROLL-SMMS2004}.

%------------------------------------------
\subsection{Discrete model}\label{disc-model}
%------------------------------------------

First, in this subsection, let us introduce a new degree of freedom
$\sigma$, which has only two-different values ($\sigma =\pm 1)$, on the triangulated lattice (see
Figure \ref{fig-A1} in  Appendix \ref{surface-model}). We  assume that
the variable $\sigma_i$ is defined on the triangle ${\it \Delta}_i$,
and moreover, the values of $\sigma_i$ correspond to two different
phases, namely, the liquid-ordered ($L_o$) and the liquid-disordered ($L_d$) phases, such that 
\begin{eqnarray}
\label{def-of-sigma} 
\sigma({\it \Delta})= \left\{
       \begin{array}{@{\,}ll}
        1 & \quad \quad \left({\it \Delta} \in L_o\right) \\
        -1 & \quad  \quad \left({\it \Delta} \in L_d\right).
       \end{array} 
       \right. 
\end{eqnarray}
This definition of $\sigma$ implies that every triangle is labelled by the value of $\sigma$, and therefore, $\sigma$ represents the phase (or domain) to which the triangle ${\it \Delta}$ belongs. 

We now introduce a discrete Hamiltonian for multi-component
membranes. The technical details of the discretization of the continuous Hamiltonian $S_1$ and $S_2$ introduced in Subsection \ref{cont-model} are described in Appendix \ref{discrete-model}, and the discrete expressions for $S_1$ and $S_2$ are given in Eq. (\ref{Disc-Eneg-model-12}) in Appendix \ref{discrete-model}. Using these $S_1$ and $S_2$, we have the total Hamiltonian $S$ such that
\begin{eqnarray}
\label{Disc-Eneg-model-2} 
&&S\left({\bf r},\sigma\right)=\lambda S_0+S_1+\kappa S_2,  \nonumber\\
&&S_0\left(\sigma\right)=\sum_{ij} \left(1-\sigma_i\cdot\sigma_j\right),  \nonumber\\
&&S_1\left({\bf r},\sigma\right)=\sum_{ij} \gamma_{ij}(\sigma) \ell_{ij}^2,\nonumber \\ 
&&S_2\left({\bf r},\sigma\right)=\sum_{ij} \kappa_{ij}(\sigma)\left(1-{\bf n}_i\cdot{\bf n}_j\right), 
\end{eqnarray}
where  $S\left({\bf r},\sigma\right)$ denotes that the Hamiltonian
depends on the variables  ${\bf r} (\in \Re^3)$ and $\sigma$. The
three-dimensional vector  ${\bf r}$ denotes the vertex position of the
triangulated lattice. The energy $\lambda S_0$ is called the {\it aggregation energy}. When $\lambda\to 0$, the variable $\sigma$ becomes random, and this random configuration simply corresponds to the coexistence phase, where $L_o$ and $L_d$ are not separated. Conversely,  when $\lambda$ becomes sufficiently large, two neighbouring $\sigma$s have the same $\sigma$, and this configuration corresponds the phases where $L_o$ and $L_d$ are separated. 
 As described above, the  second and third terms $S_1$ and $S_2$ in
 $S$ are the discrete Hamiltonians corresponding to the continuous
 ones introduced in  \ref{cont-model}. The coefficient $\kappa$ of
 $S_2$ is the bending rigidity and has units of $[1/k_BT]$, where
 $k_B$ and $T$ are  the Boltzmann constant and the temperature, respectively. In this paper, we assume that $k_BT\!=\!1$. The symbol ${\bf n}_i$ in $S_2$ expresses a unit normal vector of the triangle $i$. The symbols $\gamma_{ij}(\sigma)$ and $\kappa_{ij}(\sigma)$ denote that $\gamma_{ij}$ and $\kappa_{ij}$ depend on the variable $\sigma$, and this dependence arises from an interaction between $\sigma$ and ${\bf r}$. The interaction between $\sigma$ and ${\bf r}$ is defined by the function $\rho$ such that
\begin{eqnarray}
\label{def-of-rho} 
\rho({\it \Delta})= \left\{
       \begin{array}{@{\,}ll}
        c & \quad \quad \left({\it \Delta} \in L_o\Leftrightarrow \sigma({\it \Delta})=1\right) \\
        1 & \quad  \quad \left({\it \Delta} \in L_d\Leftrightarrow \sigma({\it \Delta})=-1\right),
       \end{array} 
       \right. 
\end{eqnarray}
where $c$ is a parameter that should be fixed at the beginning of the simulations.

%++++++++++++++++++++++++++++++++++
\begin{figure}[h]
\centering
\includegraphics[width=10.5cm]{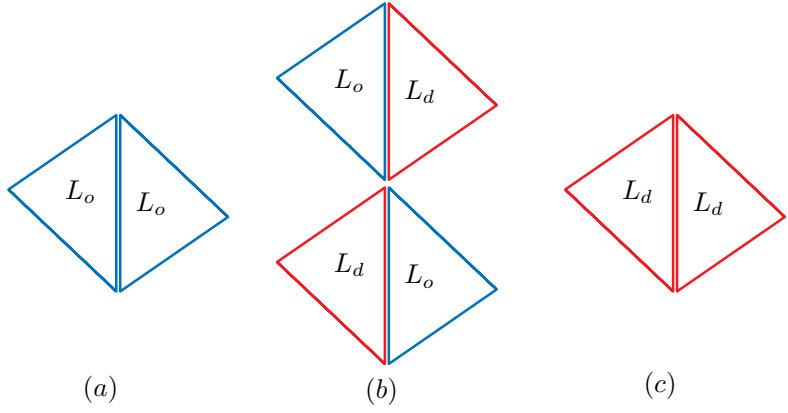}
\caption{The dependence of $\kappa_{ij}$ and $\gamma_{ij}$ on four possible combinations of $L_o$ and $L_d$:  (a) $\kappa_{ij}\!=\!\gamma_{ij}\!=\!(c\!+\!c^{-1})/2$ on $(L_o,L_o)$, (b) $\kappa_{ij}\!=\!\gamma_{ij}\!=\!(2\!+\!c\!+\!c^{-1})/4$ on $(L_o,L_d)$, and  (c) $\kappa_{ij}\!=\!\gamma_{ij}\!=\!1$ on $(L_d,L_d)$. $(L_{o,d},L_{o,d})$ correspond to the bonds represented by the duplicated lines.}
\label{fig-1}
\end{figure}
%++++++++++++++++++++++++++++++++++
 From this definition of $\rho({\it \Delta})$ and Eq. (\ref{Disc-Eneg-S-1}) in Appendix \ref{discrete-model}, we have (see Figure \ref{fig-1})
\begin{eqnarray}
\label{kappa-and-gamma} 
&&\gamma_{ij}=\kappa_{ij}=\frac{c_i+c_j}{4} =\frac{1}{4}\left(\rho_i+\frac{1}{\rho_i}+\rho_j+\frac{1}{\rho_j}\right) \nonumber \\
&&=\left\{
       \begin{array}{@{\,}ll}
        (c+c^{-1})/2 &  \quad \left[\sigma_i=\sigma_j=1\; : \;(L_o,L_o) \right] \\
        (2+c+c^{-1})/4 & \quad \left[\sigma_i\sigma_j =-1\; : \;(L_o,L_d) \right] \\
        1 & \quad \left[\sigma_i=\sigma_j=-1 \; : \;(L_d,L_d)\right].
       \end{array}        \right. 
\end{eqnarray}
These expressions represent how the effective surface tension
$\gamma_{ij}$ and bending rigidity $\kappa\kappa_{ij}$ depend on the
position of the bond $ij$, which is one of the three domain boundary
bonds $(L_o,L_o)$, $(L_o,L_d)$, and $(L_d,L_d)$. The symbol
$(L_o,L_d)$ refers to the bond shared by the two neighbouring
triangles of the $L_o$ and $L_d$ phases (see Fig. \ref{fig-1}). Note that only $(L_o,L_d)$ corresponds to the bond on the domain boundary, and the other two correspond to the bonds inside the domains $L_o$ and $L_d$. From the expressions of $\gamma_{ij}$ and $\kappa_{ij}$ in Eq. (\ref{kappa-and-gamma}), we understand that the dependence of $\gamma_{ij}$ and $\kappa_{ij}$ on the domains and their boundary is automatically determined. Thus, this expression is one of the most interesting outputs of the model in this paper.  The values of $\gamma_{ij}$ and $\kappa_{ij}$ depend on the parameter $c$, which is an input parameter. 

%++++++++++++++++++++++++++++++++++
\begin{figure}[h]
\centering
\includegraphics[width=6.5cm]{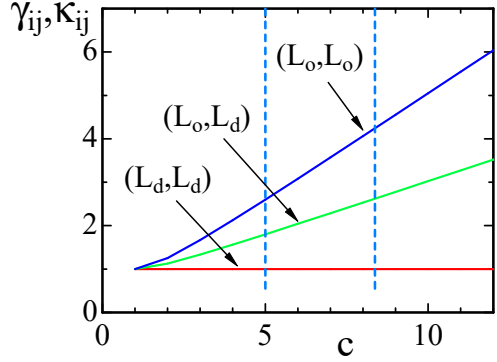}
\caption{Three different values of $\gamma_{ij}$ and $\kappa_{ij}$ vs. $c$, where $\kappa_{ij}\!=\!\gamma_{ij}\!=\!(c\!+\!c^{-1})/2$ on the $(L_o,L_o)$ boundary, $\kappa_{ij}\!=\!\gamma_{ij}\!=\!(2\!+\!c\!+\!c^{-1})/4$ on the $(L_o,L_d)$ boundary, and  $\kappa_{ij}\!=\!\gamma_{ij}\!=\!1$ on the $(L_d,L_d)$ boundary. The dashed lines denote the values of $c$ assumed in some of the simulations. }
\label{fig-2}
\end{figure}
%++++++++++++++++++++++++++++++++++
In Figure \ref{fig-2}, $\gamma_{ij}(=\!\kappa_{ij})$ for $(L_o,L_o)$,
$(L_o,L_d)$, and $(L_d,L_d)$ are plotted as functions of $c$ in the
region $1\!\leq\!c$. The expressions of $\gamma_{ij}$ and
$\kappa_{ij}$ in Eq. (\ref{kappa-and-gamma}) are symmetric under the
exchange $c\!\leftrightarrow\!1/c$, and therefore, we use the value of
$c$ rather than $1/c$ to represent $\gamma_{ij}$ and
$\kappa_{ij}$. The curve of $\gamma_{ij}(=\!\kappa_{ij})$ against $c$
is almost linear except for the region $c\!\simeq\!1$.  The dashed vertical lines in the figure correspond  to $c\!=\!5$ and $c\!=\!8.37$, which are assumed in the simulations.

The fluid surface model is defined by the sum over all possible triangulations $\sum_{\mathcal T}$ in the partition function such that
\begin{eqnarray} 
\label{Part-Func-flu}
Z(\lambda, \kappa) = \sum_{\mathcal T} \int^\prime \prod _{i=1}^{N} d {\bf r}_i \exp\left[-S({\bf r},\sigma)\right],
\end{eqnarray} 
where the prime in $\int^\prime $ denotes that the center of mass of
the surface is fixed at the origin of $\Re^3$ to protect the surface translation. The dynamical triangulation, denoted by $\sum_{\mathcal T}$, is performed using the bond flip technique  \cite{Baum-Ho-PRA1990,Ho-Baum-EPL1990,CATTERALL-PLB1989,CATTERALL-NPBSUP1992,Ambjorn-NPB1993,Noguchi-JPS2009}. Due to this bond flip, the vertices  can freely diffuse over the surface, where two neighbouring  triangles merge and split into two different ones and the total number of triangles remains unchanged in this process. Therefore, not only the vertices but also the triangles diffuse over the surface.

Note that the metric variable, or in other words, the function $\rho$, is not summed over (or integrated out) in $Z$; hence, strictly speaking, it is not a dynamical variable. However, the metric variable $\rho$ is effectively considered as dynamical in the sense that $\rho$ changes its value on the surface due to the diffusion of triangles. 

Moreover, note that the aggregation energy $\lambda S_0$ simply corresponds to the line tension energy in Refs. \cite{Yanagisawa-PRE2010,Lipowsky-SM2009}. Indeed, the energy $1\!-\!\sigma_i\cdot\sigma_j$ at the bond $ij$ has a non-zero positive value only when the bond is on the domain boundary between $L_o$ and $L_d$. More precisely, $S_0$ is proportional to the total number of bonds that form the domain boundary because the mean bond length is constant (or non-zero finite) on the boundary. 

We comment on the reason why $\lambda S_0$ is considered as the line tension energy in more detail. First, the fact that 
the mean bond length becomes constant is understood from the scale-invariant property of the partition function $Z$ in Eq. (\ref{Part-Func-flu}). Indeed, we have $\langle S_1\rangle/N\!=\!3/2$ \cite{WHEATER-JP1994}. It is easy to see that this relation is satisfied: $Z$ is independent of the scale change ${\bf r}\!\to\! \alpha {\bf r}$ for arbitrary $\alpha \in\Re$, and therefore, we have $\left. dZ(\alpha)/d\alpha \right|_{\alpha=1} =0$. Because $Z(\alpha)\!=\!\alpha^{3N-1}\sum_{\mathcal T} \int^\prime \prod _{i=1}^{N} d {\bf r}_i \exp\left[-S(\alpha {\bf r})\right]$, $S(\alpha {\bf r})\!=\!\lambda S_0\!+\!\alpha^2S_1\!+\!\kappa S_2$, we have  $S_1/N\!=\!1.5$ for sufficiently large $N$.  
The relation  $\langle S_1\rangle/N\!=\!3/2$  means that $\langle \gamma_{ij}\ell_{ij}^2\rangle$ is constant. This implies that $\langle\ell_{ij}^2\rangle$ and hence $\langle\ell_{ij}\rangle$ becomes constant. This constant $\langle\ell_{ij}\rangle$ varies depending on whether the bond $ij$ belongs to $L_o$, $L_d$ or the boundary between $L_o$ and $L_d$ because the coefficient $\gamma_{ij}$ varies depending on these domains and domain boundary as in Eq. (\ref{kappa-and-gamma}). For this reason and because $\langle\gamma_{ij}\ell_{ij}^2\rangle\!=\!{\rm constant}$, we understand that the bond length on the boundary between $L_o$ and $L_d$ becomes well defined (or non-zero finite). 
Importantly, the mean bond length is expected to be finite, although
it fluctuates around the mean value and the mean value itself varies
depending on the domains or the domain boundary.  Therefore, $\lambda
S_0$ is considered to be an extension of the line tension energy because  $S_0$ is proportional to the length of the phase boundary if the two phases are clearly separated as the domains $L_o$ and $L_d$ at least.    

The remaining problem to be clarified is how the domain boundary is formed on the triangulated surfaces. During experiments, the area fraction of $L_o$ (and $L_d$) is fixed \cite{Yanagisawa-PRE2010}. Hence, in our model the total number of triangles $N_T^o$ for $\sigma\!=\!1$ (and  $N_T^d$ for $\sigma\!=\!-1$) is fixed, where the total number of triangles 
\begin{eqnarray} 
\label{no-of-trianles}
N_T=\!N_T^o+N_T^d
\end{eqnarray} 
 is also fixed to be constant (because $N_T\!=\!2N\!-\!4$ and the total number $N$ of vertices is fixed).  
The relation between the area fraction of $L_o$ and the fraction of $N_T^o$ will be described in the next section.
Another constraint imposed on the triangles in our model is that the
value of $\sigma_i$ of triangle $i$ remains unchanged for all $i$
during the simulations. Therefore, the triangles themselves have to
diffuse over the surface to form the $L_o$ and $L_d$ domains. This triangle diffusion is numerically possible on the dynamically triangulated surfaces, which are called triangulated fluid surfaces, via the Monte Carlo (MC) technique with dynamical triangulation, as described above  \cite{Baum-Ho-PRA1990,Ho-Baum-EPL1990,CATTERALL-PLB1989,CATTERALL-NPBSUP1992,Ambjorn-NPB1993,Noguchi-JPS2009}. 
  
The function $\rho({\it \Delta})$ in Eq.(\ref{def-of-rho}) characterizes the difference between the phases $L_o$ and $L_d$ of ${\it \Delta}$, and these two different phases are labelled by the variable $\sigma({\it \Delta})$ as in Eq. (\ref{def-of-sigma}). Therefore, the model in this paper is limited to membranes with two-component domains, however, the modelling technique is  applicable to membranes with multi-component domains. Here we comment on how to extend the model to a $n$-component model. To extend the $2$-component model, we have to define  the value of $\rho({\it \Delta})$ for the $n$-component model such that   ${\it \Delta} \in L_i(1\!\leq\!i\!\leq \!n)$ (see Eq.(\ref{def-of-rho})), where $\{L_1,L_2,\cdots,L_n\}$ is the set of domains assumed. In this case, the variable  $\sigma({\it \Delta})$ should be  $n$ components, and therefore, the corresponding energy term  $\lambda S_0$ in Eq. (\ref{Disc-Eneg-model-2}) should also be extended. The Hamiltonian of $n$-states Potts model, for example, can be used for  $S_0$.  Hamiltonians of continuous models, such as the Heisenberg spin model, can also be assumed for $S_0$, where the continuous variable $\sigma({\it \Delta})$ should be connected in one-to-one correspondence with the $n$-component function $\rho({\it \Delta})$ (see Eq.(\ref{def-of-rho})). In these $n$-component models, the energy $\lambda S_0$ is still expected to play the role of line tension energy between two different domains, because $\lambda S_0$ becomes zero (nonzero) if the  phases of two neighbouring triangles are identical to (different from) each other. The parameters $\kappa_{ij}$ and $\gamma_{ij}$ are given by the same expression in Eq. (\ref{Disc-Eneg-S-1}), however, the final expression of these parameters in the $n$-component model are in general different from those in Eq. (\ref{kappa-and-gamma}) because of the dependence of the parameters on the definition of $\rho_i$. Indeed, the parameters $\kappa_{ij}$ and $\gamma_{ij}$ in the $n$-component model will be different from those determined by the single parameter $c$  in the two-component model.

%------------------------------------------
\section{Monte Carlo technique}\label{MC-technique}
%------------------------------------------
The canonical Metropolis technique is used \cite{Mepropolis-JCP-1953,Landau-PRB1976}. The vertex position ${\bf r}$ is updated such that  ${\bf r}^\prime\!=\!{\bf r}\!+\!\delta {\bf r}$. The symbol $\delta {\bf r}$ denotes a random three-dimensional vector in a sphere of radius $R$. The new position  ${\bf r}^\prime$ is accepted with probability ${\rm Min}[1, \exp(-\delta S)]$, where $\delta S\!=\!S({\rm new})\!-\!S({\rm old})$.  The radius $R$ of the small sphere is fixed such that the acceptance rate for the update of ${\bf r}$ is approximately equal to $50\%$.

The triangulation ${\mathcal T}$ is updated using the bond flip
technique, as described in the previous section
\cite{Baum-Ho-PRA1990,Ho-Baum-EPL1990,CATTERALL-PLB1989,CATTERALL-NPBSUP1992,Ambjorn-NPB1993,Noguchi-JPS2009}. We
use the same technique used in Refs.
\cite{Baum-Ho-PRA1990,Ho-Baum-EPL1990,CATTERALL-PLB1989,CATTERALL-NPBSUP1992,Ambjorn-NPB1993,Noguchi-JPS2009},
except for the following constraint. In the bond flip, the two neighbouring triangles of the bond change to a new pair of triangles such that the fraction $\phi_0$  of $L_o$ (or $L_d$) remains unchanged, where
 $\phi_0$ is defined by  
\begin{eqnarray} 
\label{fraction}
\phi_0 = N_T^o / N_T. 
\end{eqnarray} 
More precisely, if the two triangles have the same value of $\sigma$
prior to the bond flip, then the new values of $\sigma$ for the new
triangles are fixed to be the same as the old one. However, if the
values of $\sigma$ are different from each other before the bond flip,
then the new values are also fixed randomly to be different. Only
through this process is the variable $\sigma$  updated. Due to this update of $\sigma$ through the dynamical triangulation, the function $\rho$ changes, and hence, a domain structure of $L_o$ (or $L_d$) is formed on the surface. 

We comment on the relation between the fraction $\phi_0$ and the area fraction of $L_o$. As described in the previous section, the mean triangle areas $a_o$ and $a_d$ in the domains $L_o$ and $L_d$ are constant because of the scale invariance of $Z$. Therefore, the area fraction of $L_o$ can be written as $N_T^oa_o/(N_T^oa_o\!+\!N_T^da_d)$, which is identical to $\phi_0 \!=\! N_T^o / N_T$ if $a_o\!=\!a_d$. However, the area fraction of $L_o$ is not always reflected in the fraction $\phi_0$ if $a_o\!\not=\!a_d$.

The initial configuration for the simulations is fixed to be the random phase, where the $L_o$ (or $L_d$) triangles are randomly distributed on the surface under a constant ratio $\phi_0$. This random state corresponds to the two-phase coexistence configuration. 

A single Monte Carlo sweep (MCS) consists of $N$ updates of ${\bf r}$
and of $N$ updates for the bond flips.  The total number of MCS that
should be performed depends on the parameters; it ranges from
approximately $1\!\times\! 10^8$ to $8\times\!\!10^8$.  The numbers of
MCS for almost all simulations are $2\!\times\!
10^8\!\sim\!3\!\times\! 10^8$. The simulations at the phase boundaries
are relatively time consuming in general because the domain structure
and hence the surface shape changes very slowly at these boundaries. The total number of vertices $N$ is fixed to $N\!=\!5762$ in this paper.

%------------------------------------------
\section{Simulation results}\label{results}
%------------------------------------------
Two types of models, which are denoted as model 1 and model 2, are simulated.  
The Gaussian energy $S_1\!=\!\sum_{ij} \ell_{ij}^2$ of the canonical surface model is assumed for model 1. 
From this assumption, the effective surface tension $\gamma_{ij}$ for model 1 is $\gamma_{ij}\!=\!1$. Model 2 is the same as the one introduced in Section \ref{disc-model}. The Gaussian energy $S_1$ and the parameters $\gamma_{ij}$ and $\kappa_{ij}$ for model 1 and model 2 are presented in Table \ref{table-1}. 

%++++++++++++++++++++++++++++++++++
\begin{table}[h]

\caption{ The Gaussian bond potential  $S_1$ and the parameters $\gamma_{ij}$ and $\kappa_{ij}$ assumed in model 1 and model 2. (see Eq. (\ref{Disc-Eneg-S-1} ) in Appendix \ref{discrete-model}) \label{table-1}}

\begin{center}
 \begin{tabular}{|c|c|c|c|c|}
 \hline
        & $S_1$ & $\gamma_{ij}$ & $\kappa_{ij}$  \\
 \hline
model 1 & $S_1\!=\!\sum_{ij}\ell_{ij}^2$ & 1   & $(c_i+c_j)/4$             \\
 \hline
model 2  & $S_1\!=\!\sum_{ij}\gamma_{ij} \ell_{ij}^2$ &  $(c_i+c_j)/4$   &  $(c_i+c_j)/4$   \\
 \hline
 \end{tabular} 
\end{center}
 %}
\end{table}
%++++++++++++++++++++++++++++++++++
In model 1, the surface shape is influenced only by $\kappa_{ij}$
because $\gamma_{ij}\!=\!1$. In contrast, in model 2, the coefficient
$\gamma_{ij}$ influences the surface size because $S_1$ in
Eq. (\ref{Disc-Eneg-model-2}) has the unit of length squares. Indeed,
as described in Section \ref{two-comp-model}, from the scale
invariance of $Z$, we have $\langle S_1\rangle/N\!=\!3/2$
\cite{WHEATER-JP1994}, and therefore, $ \ell_{ij}^2$ deviates from the
constant expected from this relation if the constraint
$\gamma_{ij}\!=\!1$ is not imposed on $\gamma_{ij}$. For example, if
$\gamma_{ij}$ is large (small), then  $ \ell_{ij}^2$ becomes small
(large). Therefore, due to this dependence of $\ell_{ij}^2$ on
$\gamma_{ij}$, the size of the triangles in model 2 depends on the
domains. By contrast, there is no dependence of $\ell_{ij}^2$ on the
domains in model 1, where $\gamma_{ij}\!=\!1$ over the entire surface.

The input parameters for the simulations are $\lambda$, $\kappa$, $c$,
and $\phi_0$, where $c$ is the value of the function $\rho$ in
Eq. (\ref{def-of-rho}) and determines $\gamma_{ij}$ and
$\kappa_{ij}$. The  parameter $\phi_0$ defined by Eq. (\ref{fraction})
is  identical to the area fraction in model 1, whereas it differs from
the area fraction in model 2 because the triangle area is not uniform
in model 2. More precisely, the mean triangle area in the $L_o$ domain
is different from that in the $L_d$ domain in model 2. In Table \ref{table-2}, we show the parameters assumed in the simulations. The values of $\kappa_{ij}$ corresponding to the input $c$ are listed in Table \ref{table-3}.
%++++++++++++++++++++++++++++++++++
\begin{table}[h]
\caption{ The input parameters $\lambda$,  $\kappa$, $c$ and $\phi_0$ for the simulations.  \label{table-2}}
\begin{center}
 \begin{tabular}{|c|c|c|c|c|}
 \hline
  & $\lambda$ & $\kappa$  &$c$    & $\phi_0$   \\
 \hline
model 1 & $0.03\leq\lambda\leq 0.5$   & 7       & 5  &   $0.3\leq\phi_0\leq 0.8$   \\
 \hline
model 2 & $0.03\leq\lambda\leq 0.8$   & 10     & 8.37  &    $0.7\leq\phi_0\leq 0.9$  \\
model 2 & 3        & $7\leq\kappa\leq 15$      & 5  &    $0.65\leq\phi_0\leq 0.95$  \\
 \hline
 \end{tabular} 
\end{center}
\end{table}
%++++++++++++++++++++++++++++++++++
%++++++++++++++++++++++++++++++++++
\begin{table}[h]
\caption{ The input parameter  $c$ automatically defines  the values of $\kappa_{ij}$ and $\gamma_{ij}$, where $\gamma_{ij}\!=\!1$ for model 1 and $\gamma_{ij}\!=\!\kappa_{ij}$ for model 2.  \label{table-3}}
\begin{center}
 \begin{tabular}{|c|c|c|c|c|}
 \hline
  &$c$    & $\kappa_{ij}(L_o,L_o)$ & $\kappa_{ij}(L_o,L_d)$ & $\kappa_{ij}(L_d,L_d)$  \\
 \hline
model 1     & 5  & 2.6    & 1.8   & 1     \\
 \hline
model 2     & 8.37  & 4.24    & 2.62    & 1     \\
model 2     & 5  & 2.6    & 1.8    & 1     \\
 \hline
 \end{tabular} 
\end{center}
\end{table}
%++++++++++++++++++++++++++++++++++

%------------------------------------------
\subsection{Model 1}
%------------------------------------------
%++++++++++++++++++++++++++++++++++
\begin{figure}[h]
\centering
\includegraphics[width=11.5cm]{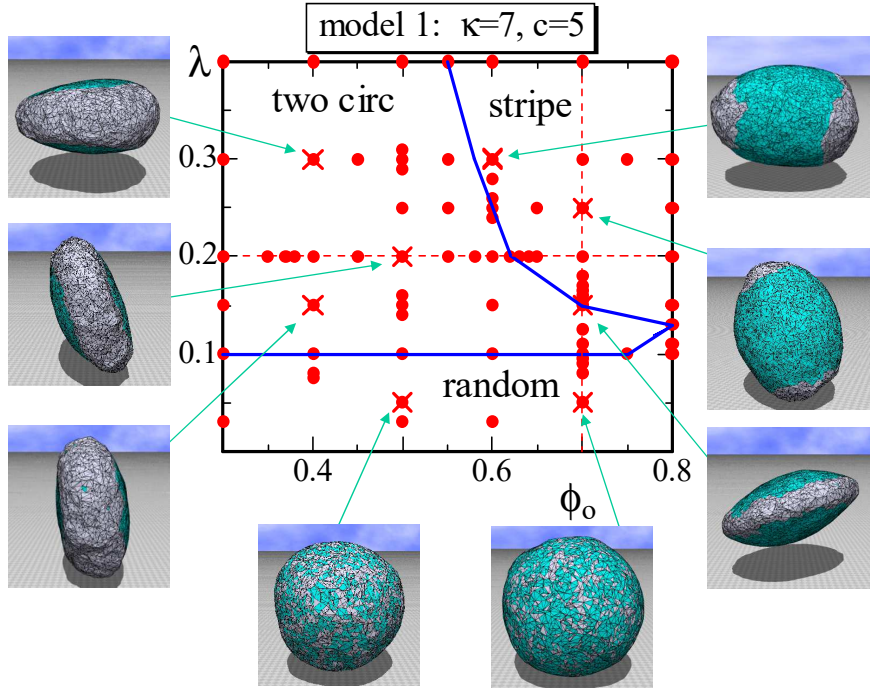}
\caption{A phase diagram of model 1 on the  $\lambda-\phi_o$ plane at $\kappa\!=\!7$ and $c\!=\!5$, and the snapshots of surfaces obtained at the points indicated by the symbol ($\textcolor{red} \times $). The solid lines denote the phase boundaries, and the dashed lines denote the positions for the simulations for Figures \ref{fig-4} (a),(b),(c).  The solid circles (\textcolor{red}{$\bullet $}) denote the data points of the simulations for the phase boundaries.  The two circular domains and the stripe domain correspond to the $L_o$ phase, which is DPPC rich. The two separated domains in the surface of striped domain and the connected domain in the surface of two circular domains correspond to the $L_d$ phase, which is DOPC rich.
  }
\label{fig-3}
\end{figure}
%++++++++++++++++++++++++++++++++++
We first show a phase diagram on the $\lambda-\phi_o$ plane in Figure
\ref{fig-3}. The parameters $\kappa$ and $c$ are fixed to
$\kappa\!=\!7$ and $c\!=\!5$, as shown in Table \ref{table-2}, and
$\lambda$ is varied in its relatively small region. The dots
(\textcolor{red}{$\bullet $}) are the data points where we perform the
simulations to construct the phase diagram.  We find that the two phases $L_o$ and $L_d$ are not
separated in the region $\lambda\!<\!0.1$, the domain pattern is
random, and the surface is almost spherical, as observed in the
snapshots.  In contrast, in the region $\lambda\!>\!0.1$, $L_o$ and
$L_d$ are clearly separated, and the two circular domains and the
stripe domain appear. The domain structure depends on the value of
$\phi_0$,  and the corresponding surface morphology appears to be
almost discontinuously separated on the phase diagram.  We observe
that the two circular domains change to the stripe domain as the
fraction $\phi_0$, which is identical with the area fraction of $L_o$,
increases for constant $\lambda$. This result is consistent with the experimental results reported in \cite{Yanagisawa-PRE2010}, where the area fraction of $L_o$ is changed. 
The two circular domains and the stripe domain correspond to the $L_o$ phase, where $\kappa_{ij}$ is higher than those of both the $L_d$ domain and the boundary, as shown in Table \ref{table-1}. For this reason, the $L_o$ domain is relatively smooth compared to the $L_d$ domain. The ratio $\kappa_{ij}(L_o,L_o)/\kappa_{ij}(L_d,L_d)(=\! 2.5\!\sim\!4.3)$ assumed in the simulations is comparable to or slightly larger than the experimental prediction $\kappa_{ij}(L_o,L_o)/\kappa_{ij}(L_d,L_d)(=\! 1\!\sim\!4)$ \cite{Yanagisawa-PRE2010}.

%++++++++++++++++++++++++++++++++++
\begin{figure}[h]
\centering
\includegraphics[width=13.5cm]{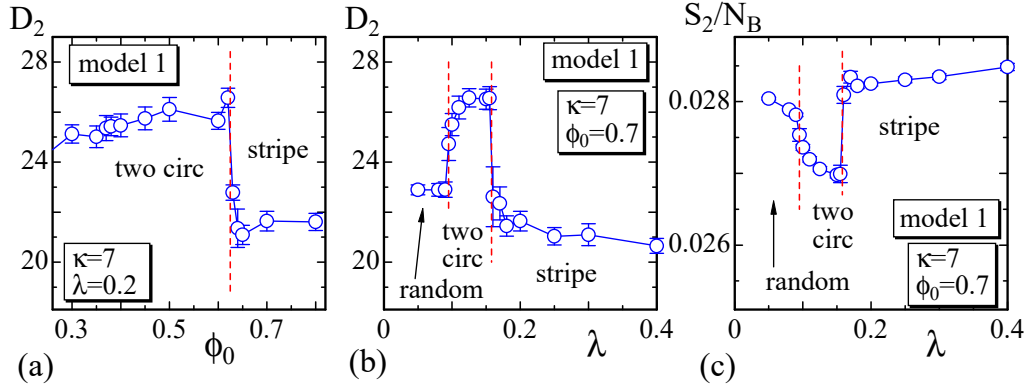}
\caption{(a) The size $D_2$ vs. $\phi_0$ at $\lambda\!=\!0.2$, (b) $D_2$ vs. $\lambda$ at $\phi_0\!=\!0.7$, and (c) the bending energy $S_2/N_B$ vs. $\lambda$ at $\phi\!=\!0.7$. These are calculated  on the dashed horizontal and vertical lines in Figure \ref{fig-3}. The minor axis $D_2$ and the bending energy  $S_2/N_B$  change almost discontinuously and smoothly at the phase boundaries, which are denoted by the vertical dashed lines. }
\label{fig-4}
\end{figure}
\begin{figure}[h]
\centering
\includegraphics[width=7.5cm]{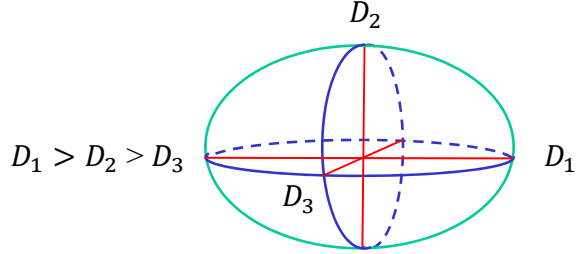}
\caption{ The surface size is characterized by three diameters $D_1$, $D_2$, and $D_3$, where $D_1\!>\!D_2\!>\!D_3$. The three axes are perpendicular to each other.}
\label{fig-5}
\end{figure}

%++++++++++++++++++++++++++++++++++
Next, to show the dependence of the surface size on the parameters, we
define semi-axis lengths  $D_1$, $D_2$, and $D_3$ of the surface such
that $D_1\!>\!D_2\!>\!D_3$ as in Figure \ref{fig-5}. $D_1$ and $D_2$,
$D_3$  correspond to the major and minor axes, respectively. The
surface of the stripe domain corresponds to the so-called prolates,
where $D_1>D_2\simeq D_3$ is expected. It is also expected  that
$D_1\simeq D_2>D_3$ in the so-called oblates, which corresponds to the
surface shape of the two circular domains. Therefore, the surfaces
with the stripe and two circular domains can be distinguished by the minor axis $D_2$.   

%++++++++++++++++++++++++++++++++++
%++++++++++++++++++++++++++++++++++

We plot $D_2$ vs. $\phi_0$ in Figure \ref{fig-4}(a), where $\lambda\!=\!0.2$.
As shown, $D_2$ discontinuously changes against  $\phi_0$ at the phase
boundary between the two circular and stripe domains. From the plot of
$D_2$ vs. $\lambda$ in  Figure \ref{fig-4}(b), we also observe that
$D_2$ discontinuously changes at the same phase boundary. The bending
energy $S_2/N_B$ in Figure \ref{fig-4}(c) also discontinuously changes
at this boundary, and this result indicates that this morphological
change is considered as a first-order transition. However,  note that
the change of the morphology at this phase boundary is relatively smooth. 
In fact, one circular domain surface, which is not shown as a
snapshot in Figure  \ref{fig-3}, can be observed at the boundary. This
implies that the stripe domain surface and one circular domain surface
have the same bending energy, or in other words, the bending energy is
degenerate.  Additionally, note that the phase boundary between the
two circular and random domains appears to be continuous. This means
that the shape of the two circular domain surface continuously changes to the random domain surface. At this phase boundary, the surface shape continuously changes from pancake to sphere.  

%------------------------------------------
\subsection{Model 2}
%------------------------------------------
%++++++++++++++++++++++++++++++++++
\begin{figure}[h]
\centering
\includegraphics[width=11.5cm]{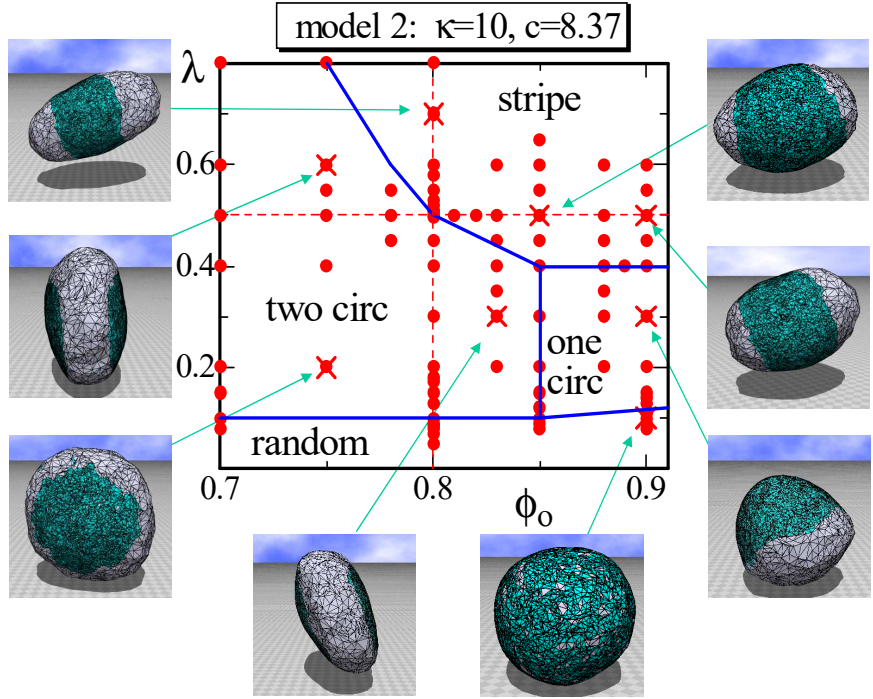}
\caption{A phase diagram of model 2 on the  $\lambda-\phi_o$ plane at $\kappa\!=\!10$ and $c\!=\!8.37$ and the snapshots of surfaces. The solid lines on the phase diagram denote the phase boundaries, and the dashed lines denote the positions for the simulations for Figures \ref{fig-7} (a),(b),(c). The solid circles (\textcolor{red}{$\bullet $}) denote the data points of the simulations for the phase boundaries.} 
\label{fig-6}
\end{figure}
%++++++++++++++++++++++++++++++++++
In model 2, not only $\kappa_{ij}$ but also $\gamma_{ij}$ depend on
the domain (or the domain boundary) whether it is $L_o$ or $L_d$. For
this reason, the area of the triangles in the $L_o$ domain becomes
considerably smaller than that in the $L_d$ domain. Therefore, the
fraction $\phi_0$ does not reflect the area fraction of $L_o$ in this
case. In fact, it is easy to see that the area fraction of $L_o$ in
the snapshots at $\phi_0\!=\!0.9$ in Figure \ref{fig-6} is much
smaller than $90\%$. Nevertheless, the phase diagram on the
$\lambda-\phi_0$ plane  in Figure \ref{fig-6} appears almost the same
as that in Figure \ref{fig-3}. The only difference between the two
phase diagrams is the appearance of one circular domain phase, denoted
by "one circ" in Figure \ref{fig-6}. This one circular phase is
stable, where "stable" means that the surface domain remains unchanged
against a small variation of the parameters inside the phase
boundary. This is in sharp contrast to the one circular domain
surfaces observed at the region close to the boundary between the
two circular and stripe domains because these one circular surfaces
are very sensitive to the parameter variation and hence
"unstable". The shape of the one circular surface in the one circular
region  is almost spherical, such as the one shown in Figure
\ref{fig-6}, and this result is in contrast to the result in
Ref. \cite{Lipowsky-SM2009}, where the one circular phase is separated
into two phases: the prolate and oblate phases. One possible reason for
why only a spherical surface appears in the one circular domain in Figure \ref{fig-6} is because the $L_o$ domain is hardly bent due to the high ratio $\kappa_{ij}(L_o,L_o)/\kappa_{ij}(L_d,L_d)\!=\! 4.24$, which is slightly larger than the one $1\!<\!\kappa_{ij}(L_o,L_o)/\kappa_{ij}(L_d,L_d)\!<\!3$ assumed  in  Ref. \cite{Lipowsky-SM2009}. 
The  parameters assumed on this plane are $\kappa\!=\!10$ and $c\!=\!8.37$, which are listed in Table \ref{table-1}.

The simulations are also performed on the $\lambda-\phi_0$ planes for
larger $\kappa$, such as $\kappa\!=\!15$ and  $\kappa\!=\!20$, and
with $c\!=\!8.37$.  The phase diagrams obtained in these simulations
are (not shown) relatively close to that shown in Figure \ref{fig-6};
however, surfaces with three or four circular domains appear in the
lower $\lambda$ region in the two circular domain phase. The bending
energy $\kappa S_2$ of the three or four domains is lower than that of
the two circular domain; moreover, the aggregation energy $\lambda
S_0$ of these multi-circular domains is larger than that of the two circular domain. These are the reasons for the appearance of the three or four domains only in the relatively small $\lambda$ region in the simulations with relatively large $\kappa$.

%++++++++++++++++++++++++++++++++++
\begin{figure}[h]
\centering
\includegraphics[width=13.5cm]{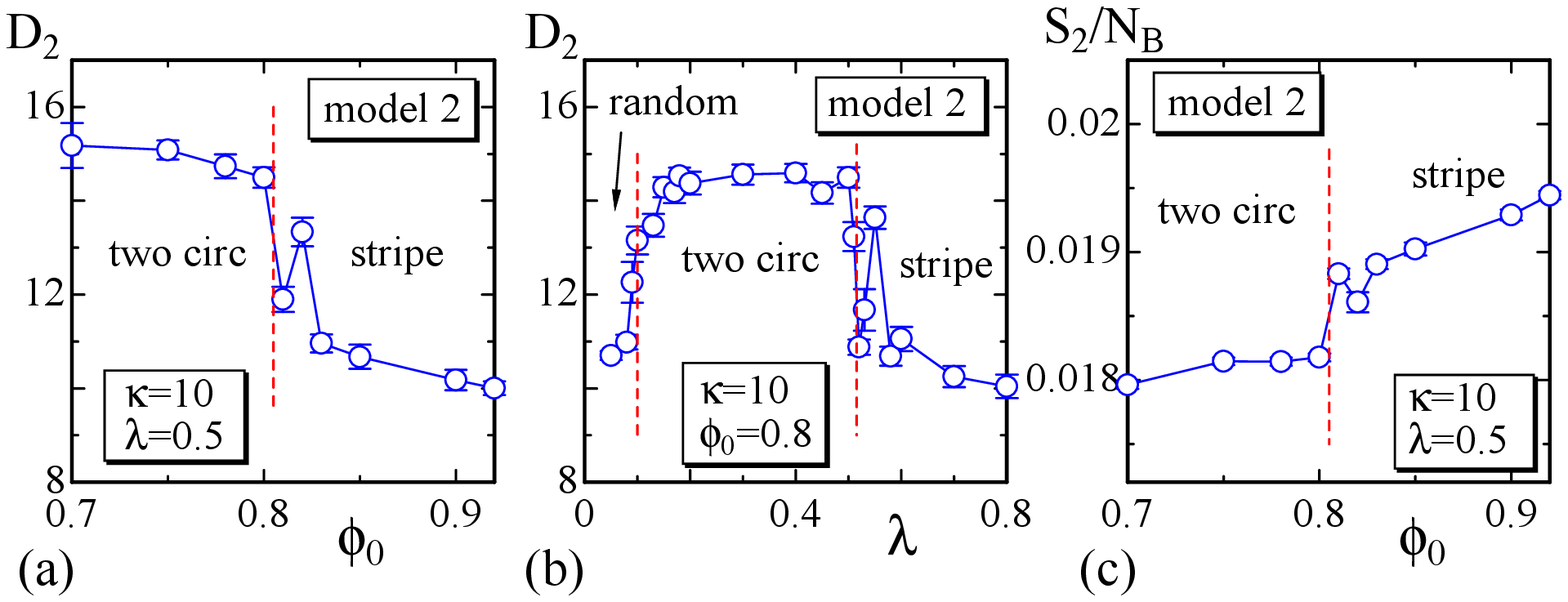}
\caption{(a) The size $D_2$ vs. $\phi_0$ at $\lambda\!=\!0.5$, (b) $D_2$ vs. $\lambda$ at $\phi_0\!=\!0.8$, and (c) the bending energy $S_2/N_B$ vs. $\phi_0$ at $\lambda\!=\!0.5$. These are calculated  on the dashed horizontal and vertical lines in Figure \ref{fig-6}. The size of the surface changes almost discontinuously and smoothly at the phase boundaries, which are denoted by the dashed lines.}
\label{fig-7}
\end{figure}
%++++++++++++++++++++++++++++++++++
To observe the variation of the surface size at the phase boundaries,
we calculate the size $D_2$ on the dashed lines in Figure \ref{fig-6}
and plot them in Figures \ref{fig-7}(a), (b). We determine that $D_2$
discontinuously changes against $\phi_0$ and $\lambda$ at the phase
boundaries, similar to that in model 1 shown in Figures
\ref{fig-4}(a), (b). Moreover, the phase boundary is also not as clear
because of the same reason as that for model 1. In fact, the surface
shape at the phase boundary between the two circular and stripe
domains is not always stable in model 2, similar to that in  model
1. Figure \ref{fig-7}(c) also shows that $S_2/N_B$ discontinuously
changes; however, the gap is very small, and these two phases are
hence separated by a weak first-order transition. The phase boundary
between the two circular and the random domains is also expected to be continuous in  model 2. The boundaries of one circular to two circulars and one circular to stripe are also not as clear, and the boundary of one circular to random is continuous. 

%++++++++++++++++++++++++++++++++++
\begin{figure}[h]
\centering
\includegraphics[width=11.5cm]{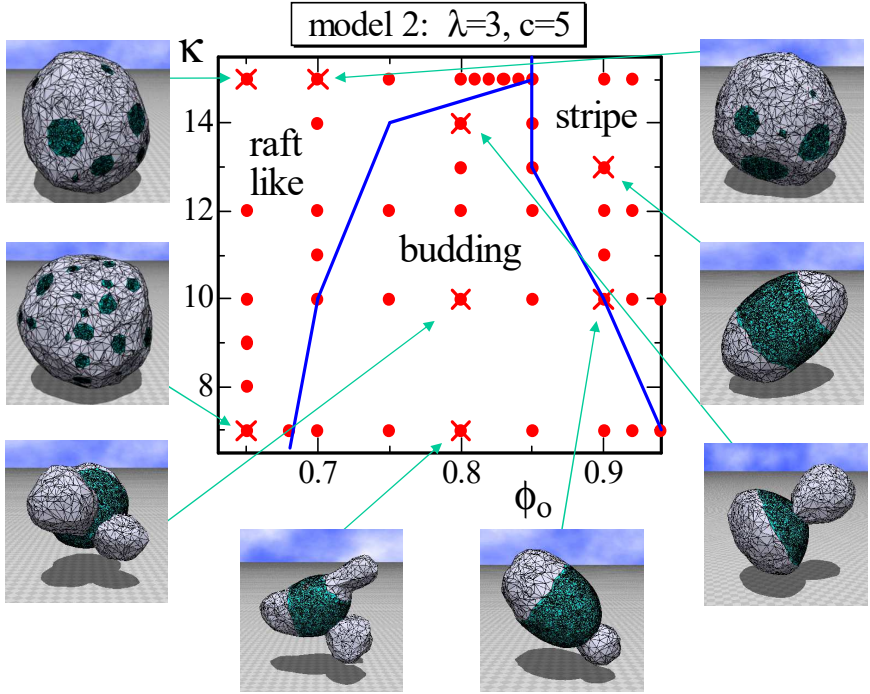}
\caption{A phase diagram of model 2 on the  $\kappa\!-\!\phi_o$ plane at $\lambda\!=\!3$ and $c\!=\!5$ and the snapshots of surfaces. The solid lines on the phase diagram denote the phase boundaries. The solid circles (\textcolor{red}{$\bullet $}) denote the data points of the simulations for the phase boundaries.} 
\label{fig-8}
\end{figure}
%++++++++++++++++++++++++++++++++++
Another difference between model 1 and model 2, other than the
appearance of the stable one circular domain,  is the raft-like domain
and the budding domain. More precisely, the budding domain can also be
seen in model 1; however, it is more clear in model 2.  The phase
diagram of model 2 on the $\kappa\!-\!\phi_0$ plane is drawn in Figure
\ref{fig-8}. The parameter $\lambda$ is fixed to  $\lambda\!=\!3$,
which is relatively large compared with the previous one assumed in
the simulations for Figures \ref{fig-3} and
\ref{fig-6}. Consequently, the energy $\lambda S_0$, which is the line
tension energy, at the phase boundary between $L_0$ and $L_d$ becomes
large in the region where $\kappa$ is relatively small. This is the
reason why the budding domain appears on this $\kappa\!-\!\phi_0$
plane in Figure \ref{fig-8}. Note that the budding domain in some of
the budding surfaces goes inside the surface and some of them
self-intersect because no self-avoiding interaction is assumed. Figure
\ref{fig-8} also shows that the raft-like domain is stable in the
relatively large $\kappa$ region, where the surface hardly
deforms. The reason why the raft domains, which are multi-circular
domains, appear only at the region of small $\phi_0$ is because the
multi-circular domains are more energetically favourable than the
stripe domain. Indeed, the effective bending rigidity
$\kappa\kappa_{ij}$ and hence $\kappa S_2$ become very large on the
large connected $L_o$ domain, such as the stripe domain, where the
line tension energy $\lambda S_0$ is relatively small. Note that $S_0$
has a non-zero positive value only on the boundary bonds between $L_o$
and $L_d$, while $S_2$ has a non-zero value on all of the
bonds. Moreover, note that the boundary length between $L_o$ and $L_d$ becomes longer (shorter) if the total number of circular domains increases (decreases), whereas the areas of $L_o$ and $L_d$ remain constant and are independent of the total number of $L_o$ domains.

%++++++++++++++++++++++++++++++++++
\begin{figure}[h]
\centering
\includegraphics[width=10.5cm]{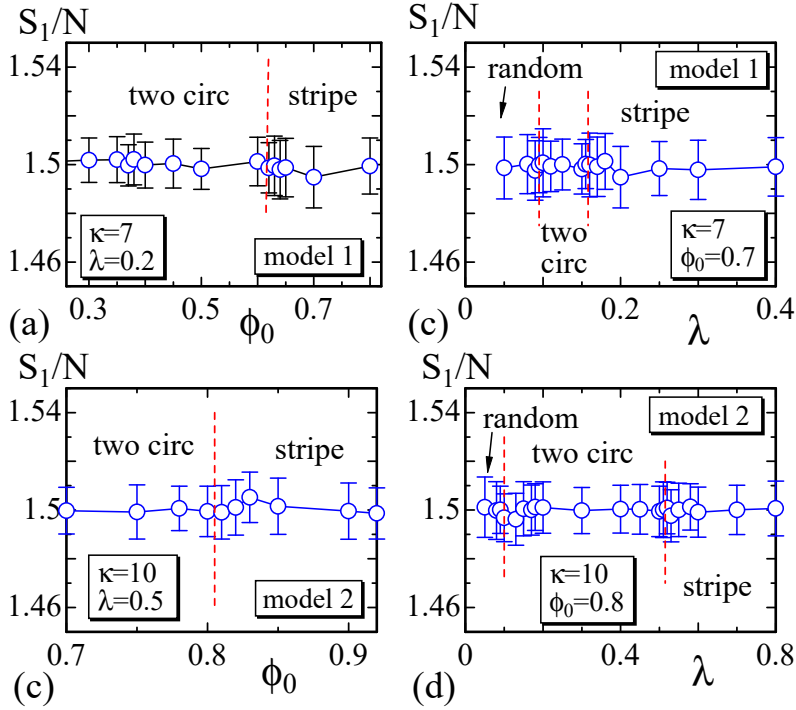}
\caption{ (a) The Gaussian energy $S_1/N$ vs. $\phi_0$ at $\lambda\!=\!0.2$,  (b) $S_1/N$  vs. $\lambda$ at $\phi_0\!=\!0.7$ for model 1, (c) $S_1/N$  vs. $\phi_0$ at  $\lambda\!=\!0.5$, and (d)  $S_1/N$  vs. $\lambda$ at $\phi_0\!=\!0.8$ for model 2. The data in (a) and (b) ((c) and (d)) are obtained on the dashed lines in Figure \ref{fig-3} (Figure \ref{fig-6}). }
\label{fig-9}
\end{figure}
%++++++++++++++++++++++++++++++++++
Finally, we show that $S_1/N$ satisfies the relation $S_1/N\!=\!1.5$, which is expected by the scale invariance of $Z$ in Eq. (\ref{Part-Func-flu}) \cite{WHEATER-JP1994}. As described in Section \ref{two-comp-model}, the bond length is expected to be well defined in the sense that the mean bond length is constant on the surface, although this constant varies depending on the domains or the domain boundary to which the bond belongs. The data in Figures \ref{fig-9}(a), (b) are obtained on the dashed lines in Figure \ref{fig-3}, and those in  Figures \ref{fig-9}(c), (d) are obtained on the lines in Figure \ref{fig-6}. These data shown in Figure \ref{fig-9}  indicate that the simulations including the energy discretization are successful.  

%------------------------------------------
\section{Summary and conclusion}\label{conclusion}
%------------------------------------------
We have studied the phase separation of the three-component membrane
with DPPC, DOPC, and cholesterol using a Finsler geometry (FG) surface
model.  The FG model is obtained from the Helfrich-Polyakov (HP) model
for membranes by replacing the surface metric with a general one
$g_{ab}\!\not=\!\delta_{ab}$, which can be called the Finsler
metric. In other words, we have extended the HP model to
 explain the morphological changes of the three-component membranes
 in the context of FG modelling.
 This new model includes a new degree of freedom $\sigma$,
which represents the liquid-ordered ($L_o$) and liquid-disordered
($L_d$) domains. The results obtained from Monte Carlo (MC)
simulations are consistent with the experimental results that have
been reported in the literature. We confirm the phase separation
of the $L_o$ and $L_d$ domains on the surface and that the surface
shows a variety of morphologies, such as the two circular domain,  the
stripe domain, the raft domain, and the budding domain.

The line tension energy, which has been used for understanding the morphological changes, simply  corresponds to the aggregation energy term $\lambda S_0$ in our model. Indeed, the value of $S_0$ is only the total number of bonds on the boundary between $L_o$ and $L_d$ in our new model. Moreover, the fact that {$\lambda S_0$} is simply the line tension energy implies that the line tension originates from the interaction between the domains because the interaction between the variables $\sigma$ in $S_0$ describes the interaction between the domains.  This interaction is closely connected to the property of the new model that the surface strength, such as the surface tension and the bending rigidity, is dependent on the bond position on the surface. This property arises from the interaction between $\sigma$ and ${\bf r}$ introduced via the Finsler metric. 

%%%%%%%%%%%%%%%%%%%%%%%%%%%%%%%%%%%%%%%%%%
\vspace{6pt} 

%%%%%%%%%%%%%%%%%%%%%%%%%%%%%%%%%%%%%%%%%%
%% optional
%\supplementary{The following are available online at www.mdpi.com/link, Figure S1: title, Table S1: title, Video S1: title.}

%%%%%%%%%%%%%%%%%%%%%%%%%%%%%%%%%%%%%%%%%%
%\acknowledgments{
We acknowledge Hideo Sekino, Andrey Shobukhov,  Giancarlo Jug and Andrei Maximov for discussions. 
This work is supported in part by JSPS KAKENNHI Number 26390138. 
%}

%%%%%%%%%%%%%%%%%%%%%%%%%%%%%%%%%%%%%%%%%%

%%%%%%%%%%%%%%%%%%%%%%%%%%%%%%%%%%%%%%%%%%

%%%%%%%%%%%%%%%%%%%%%%%%%%%%%%%%%%%%%%%%%%
%% optional
\appendix
%------------------------------------------
\section{Finsler geometry modelling}\label{surface-model}
%------------------------------------------
%------------------------------------------
\subsection{Discrete surface model}\label{discrete-model}
%------------------------------------------
%++++++++++++++++++++++++++++++++++
\begin{figure}[h]
\centering
\includegraphics[width=13.5cm]{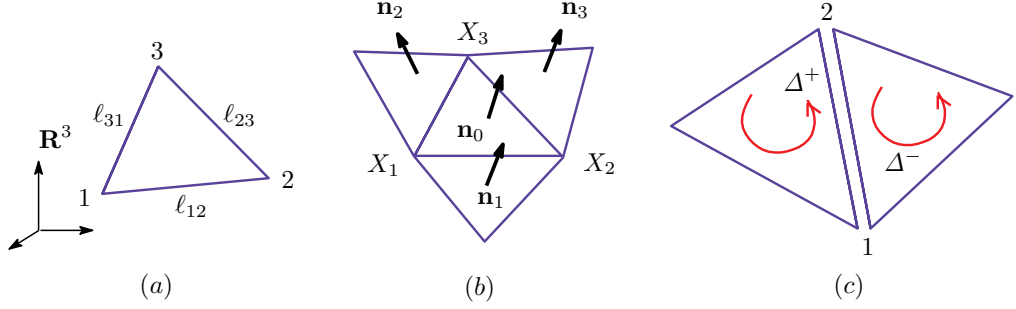}
\caption{(a) A triangle ${\it \Delta}$ included in a triangulated
  sphere in $\Re^3$; (b) the three nearest neighbour triangles of
  ${\it \Delta}$ and the unit normal vectors ${\bf n}_0$, ${\bf n}_1$,
  ${\bf n}_2$ and ${\bf n}_3$; and (c) the triangle orientation that defines the direction-dependent bond potential $\gamma_{12}\ell_{12}^2$ and $\gamma_{21}\ell_{21}^2$ of the bond $12$, where $\ell_{12}\!=\!\ell_{21}$.  }
\label{fig-A1}
\end{figure}
%++++++++++++++++++++++++++++++++++

To obtain the discrete model from the continuous surface model introduced in Section \ref{cont-model}, we assume that the surface is triangulated in $\Re^3$. The Hamiltonian is defined on  the triangulated surfaces, which are composed of three simplexes such as vertices, bonds, and triangles. Thus, all physical quantities, including Hamiltonians and the metric function, are defined on these simplexes labelled by integers. For example, the vertex position ${\bf r}_i$ is defined at the vertex $i$, the bond length $\ell_{ij}$ is defined on the bond $ij$, and the elements of $g_{ab}$ are defined on the triangle  ${\it \Delta}$. Note that the variable ${\bf r}$ is considered as a mapping from the parameter space $M$ to $\Re^3$.

We start with the discrete metric $g_{ab}$ such that
\begin{equation}
\label{metric}
g_{ab}=\left(  
       \begin{array}{@{\,}ll}
        1/\rho & \; 0 \\
        0 & \; \rho 
       \end{array} 
       \\ 
 \right), \quad \rho>0,
\end{equation} 
where $\rho$ is a function on a triangle ${\it \Delta}(\subset \Re^3)$ (see Figure \ref{fig-A1}(a)). More precisely, the elements of $g_{ab}$ are functions on the triangle ${\it \Delta}_M (\subset M)$, where $M$ is the aforementioned two-dimensional space $M$ (independent of $\Re^3$).  We assume that $M$ is also triangulated by the triangles  ${\it \Delta}_M$.  On this ${\it \Delta}_M$, an orthogonal coordinate can be taken for any one of three vertices \cite{FDAVID-SMMS2004}. For this reason, the metric $g_{ab}$ can be diagonalizable.
The inequality $\rho\!>\!0$ in Eq. (\ref{metric}) is necessary for the
positivity of the bond length. This metric depends only on $x$ and is
independent of $y$, and hence, it simply corresponds to the Riemannian metric. Indeed, this metric in Eq. (\ref{metric}) comes from the most general one, such as $g_{ab}\!=\!\left(  
       \begin{array}{@{\,}ll}
        E & \; F \\
        F & \; G 
       \end{array} 
       \\ 
 \right)$, with the functions of $E\!>\!0,\; G\!>\!0, EG\!-\!F^2\!>\!0$. By assuming $F\!=\!0$, we have $g_{ab}\!=\!\left(  
       \begin{array}{@{\,}ll}
        E & \; 0 \\
        0 & \; G 
       \end{array} 
       \\  \right) \!=\! E\left(  
       \begin{array}{@{\,}ll}
        1 & \; 0 \\
        0 & \; G/E 
       \end{array} 
       \\  \right)\!\simeq\! \left(  
       \begin{array}{@{\,}ll}
        1 & \; 0 \\
        0 & \; \rho^2 
       \end{array} 
       \\  \right)\!\simeq\! \left(  
       \begin{array}{@{\,}ll}
        1/\rho & \; 0 \\
        0 & \; \rho 
       \end{array} 
       \\  \right)$, where  $\rho^2\!=\!G/E$ and the symbol "$\simeq$"
       denotes conformally equivalent. Note that this expression of
       $g_{ab}$ depends on the local coordinates on ${\it \Delta}$,
       and therefore, the expression of $g_{ab}$ implicitly depends on
       the vertex of ${\it \Delta}$ because the coordinate origin is
       located on one of the vertices of ${\it \Delta}$. In the
       discrete models that have been studied thus far, the Euclidean metric $g_{ab}\!=\!\delta_{ab}$ (or the induced metric $g_{ab}\!=\!\partial_a {\bf r}\cdot\partial_b {\bf r}$) is always assumed as mentioned above, and it has been reported that the model for polymerized membranes undergoes a discontinuous or a continuous transition between the crumpled phase and the smooth phase \cite{KD-PRE2002,Kownacki-Mouhanna-2009PRE,Essa-Kow-Mouh-PRE2014,Cuerno-etal-2016}. 

Let the vertex ${\bf r}_1$ of the central triangle ${\it \Delta}$ in Figure \ref{fig-A1}(b) be the local coordinate origin in this ${\it \Delta}$.  By replacing 
\begin{eqnarray}
\label{replace}
&&\int \sqrt{g}d^2x \to \sum_{\it \Delta}, \nonumber \\
&&\frac{\partial {\bf r}}{\partial x_1} \to  {\bf r}_2 - {\bf r}_1, \quad  \frac{\partial {\bf r}} {\partial x_2}\to  {\bf r}_3-{\bf r}_1, \nonumber \\
&&\frac{\partial {\bf n}}{\partial x_1} \to  {\bf n}_0 - {\bf n}_2, \quad \frac{\partial {\bf n}}{\partial x_2} \to  {\bf n}_0-{\bf n}_1,
\end{eqnarray}
we have 
\begin{eqnarray}
\label{Disc-Eneg-model-10}
&&S_1=\sum_{\it \Delta}S_1\left({\it \Delta}\right)=\sum_{\it \Delta}\left(\rho \ell_{12}^2+\frac{1}{\rho}\ell_{13}^2\right),\nonumber \\
&&S_2=\sum_{\it \Delta}S_2\left({\it \Delta}\right)=\sum_{\it \Delta} \left[\rho\left(1-{\bf n}_0\cdot{\bf n}_1\right)+\frac{1}{\rho} \left(1-{\bf n}_0\cdot{\bf n}_2\right)\right], 
\end{eqnarray}
where ${\bf n}_i(i\!=\!1,2,3)$ are the unit normal vectors shown in Figure \ref{fig-A1}(b). The symbol $\ell_{ij}(=\ell_{ji})$ is defined by $\ell_{ij}\!=\!|{\bf r}_j\!-\!{\bf r}_i|$.  Note that the unit normal vector also represents the surface orientation; indeed, ${\bf n}_0$ is defined by ${\bf n}_0\!=\!{\vec \ell}_{12}\times{\vec \ell}_{13}/|{\vec \ell}_{12}\times{\vec \ell}_{13}|$ for example.

We have three possible coordinate origins in the triangles. For this
reason, $S_1$ and $S_2$ can be symmetrized by including the terms that
are cyclic permutations, such as $1\!\to\! 2$, $2\!\to\! 3$, $3\!\to\!
1$ for  $\ell_{ij}$, ${\bf n}_i$ and $\rho_i$. Summing over all
possible terms and multiplying by a factor of $1/3$, we obtain 
\begin{eqnarray}
\label{Disc-Eneg-model-11}
S_1=&& \frac{1}{3}\sum_{\it \Delta} \left[\left(\rho_{1}+\frac{1}{\rho_{2}}\right) \ell_{12}^2+\left(\rho_{2}+\frac{1}{\rho_{3}}\right)\ell_{23}^2+\left(\rho_{3}+\frac{1}{\rho_{1}}\right)\ell_{31}^2\right], \nonumber \\
S_2=&& \frac{1}{3}\sum_{\it \Delta} \left[\left(\rho_{1}+\frac{1}{\rho_{2}}\right) \left(1-{\bf n}_0\cdot{\bf n}_1\right)+\left(\rho_{2}+\frac{1}{\rho_{3}}\right)\left(1-{\bf n}_0\cdot{\bf n}_3\right) \right.\nonumber\\ 
&&\qquad +\left.\left(\rho_{3} 
+\frac{1}{\rho_{1}}\right)\left(1-{\bf n}_0\cdot{\bf n}_2\right)\right],
\end{eqnarray}
where $\rho_{i}(i\!=\!1,2,3)$ are defined on the triangle ${\it
  \Delta}$. The reason for why these three different functions $\rho_{i}$ are included is because the expression for $g_{ab}$ generally depends on the local coordinate as mentioned above.  More precisely, $\rho_{i}$ is the element of  $g_{ab}$  on ${\it \Delta}$ where the coordinate origin is located at vertex $i$. For arbitrary  $g_{ab}$, we always have the metric of the form in Eq. (\ref{metric})  by the same procedure as described above.  
 
Here, we further simplify the model by assuming that 
\begin{eqnarray}
\label{assumption} 
\rho_{1}=\rho_{2}=\rho_{3}(=\rho_{\it \Delta}).
\end{eqnarray}
Thus, we have the expressions 
\begin{eqnarray}
\label{Disc-Eneg-model-11-1}
&&S_1=\frac{1}{3}\sum_{\it \Delta} \left(\rho_{\it \Delta}+\frac{1}{\rho_{\it \Delta}}\right) \left(\ell_{12}^2+\ell_{23}^2+\ell_{31}^2\right),\nonumber \\
&&S_2=\frac{1}{3}\sum_{\it \Delta} \left(\rho_{\it \Delta}+\frac{1}{\rho_{\it \Delta}}\right) \left(1-{\bf n}_0\cdot{\bf n}_1+ 1-{\bf n}_0\cdot{\bf n}_2+1-{\bf n}_0\cdot{\bf n}_3\right).
\end{eqnarray}
Replacing the sum over triangles $\sum_{\it \Delta}$ with the sum over bonds $\sum_{ij}$, we obtain
\begin{eqnarray}
\label{Disc-Eneg-model-11-2}
 &&S_1=\frac{1}{3}\sum_{ij} \left(\rho_i+\frac{1}{\rho_i}+\rho_j+\frac{1}{\rho_j}\right) \ell_{ij}^2,\nonumber \\ 
&&S_2=\frac{1}{3}\sum_{ij} \left(\rho_i+\frac{1}{\rho_i}+\rho_j+\frac{1}{\rho_j}\right) \left(1-{\bf n}_i\cdot{\bf n}_j\right). 
\end{eqnarray}
Note that $S_1$ and $S_2$ defined by the sum over triangles $\sum_{\it
  \Delta}$ in Eq. (\ref{Disc-Eneg-model-11-1}) are exactly same as
those defined by the sum over bonds $\sum_{ij}$ in
Eq. (\ref{Disc-Eneg-model-11-2}), and the difference is only in their
expressions.  Additionally, note that the suffixes $i,j$ of $\ell_{ij}$ in Eq. (\ref{Disc-Eneg-model-11-2}) denote the bond $ij$, whereas those of $\rho_i$ and  $\rho_j$ denote the two neighbouring triangles $i$ and $j$ of the bond $ij$.  
Thus, we finally have 
\begin{eqnarray}
\label{Disc-Eneg-model-12} 
&&S\left({\bf r},\sigma\right)=S_1+\kappa S_2,  \nonumber\\
&&S_1=\sum_{ij} \gamma_{ij} \ell_{ij}^2,\quad
S_2=\sum_{ij} \kappa_{ij}\left(1-{\bf n}_i\cdot{\bf n}_j\right), 
\end{eqnarray}
with
\begin{eqnarray}
\label{Disc-Eneg-S-1} 
\gamma_{ij}= \kappa_{ij}= \frac{c_i+c_j}{4}, \quad 
c_i=\rho_i+\frac{1}{\rho_i},
\end{eqnarray}
where the irrelevant numerical factor $1/3$ is replaced by $1/4$ in the final expressions for $S_1$ and $S_2$. 

Note that $\gamma_{ij}$ ($\kappa\kappa_{ij}$) can be called the effective surface tension (effective bending rigidity) on the bond between vertices $i$ and $j$. It must be emphasized that the quantities $\gamma_{ij}$ and $\kappa_{ij}$ are independent of the bond direction, or in other words, these are symmetric under the exchange of $i$ and $j$, and for this reason, $\gamma_{ij}$ and $\kappa_{ij}$ are considered as the quantities defined on the bond $ij$. Indeed, from the expression given in Eq. (\ref{Disc-Eneg-S-1}), we have 
\begin{eqnarray}
\label{symmetry} 
\gamma_{ij}=\gamma_{ji},\quad \kappa_{ij}=\kappa_{ji}.
\end{eqnarray}
Therefore, the physical quantities $\gamma_{ij}\ell_{ij}^2$ in $S_1$ and $\kappa_{ij}(1\!-\!{\bf n}_i\cdot{\bf n}_j)$ in $S_2$ of Eq. (\ref{Disc-Eneg-model-12}) are well defined in the sense that these quantities are symmetric under the exchange of $ij$. The reason why we need this symmetry in the physical quantities $\gamma_{ij}\ell_{ij}^2$ and $\kappa_{ij}(1\!-\!{\bf n}_i\cdot{\bf n}_j)$ is because these quantities correspond to the energies for the expansion and bending of the surface at the bond $ij$, and these energies are independent of the bond direction such as the one from $i$ to $j$ or the reverse.  
 Thus, the symmetry property in Eq. (\ref{symmetry}) allows us to call
 $\gamma_{ij}$ and $\kappa\kappa_{ij}$ the effective surface tension
 and the effective bending rigidity on the bond $ij$,
 respectively. However, as we will show in the next subsection,
 $\gamma_{ij}$ and $\kappa_{ij}$ are not symmetric in general (see
 Figure \ref{fig-A1}(c)).  Moreover, note that this problem of whether $\gamma_{ij}$ and $\kappa_{ij}$ are symmetric arises only when $\gamma_{ij}$ and $\kappa_{ij}$ depend on the functions $\rho_i$ and $\rho_j$ on the two neighbouring triangles $i$ and $j$.  This is in sharp contrast to the case where $\gamma_{ij}$ and $\kappa_{ij}$ depend only on the quantity defined on the vertices \cite{Koibuchi-Sekino-2014PhysicaA}, where $\gamma_{ij}$ and $\kappa_{ij}$ are always symmetric. This exchange symmetry/asymmetry reflects the orientation symmetry/asymmetry, which will be discussed in the next subsection. 

%------------------------------------------
\subsection{Finsler geometry model}\label{surface-orientation}
%------------------------------------------
%++++++++++++++++++++++++++++++++++
\begin{figure}[h]
\centering
\includegraphics[width=13.5cm]{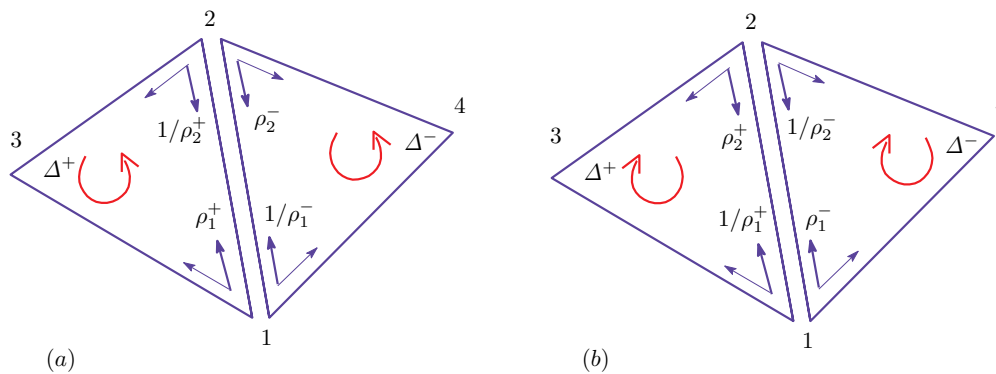}
\caption{Local coordinate origins of the triangles ${\it \Delta}^+$ and ${\it \Delta}^-$ for $S_1(\ell_{12})$, and the elements of $\gamma_{12}$ and  $\gamma_{21}$  of the configurations of (a) the original and (b) the inside out (inside view). }
\label{fig-A2}
\end{figure}
%++++++++++++++++++++++++++++++++++
In this subsection, we show that the discrete surface models constructed above are well defined only in the context of Finsler geometry modelling \cite{Koibuchi-Sekino-2014PhysicaA}. For this purpose, we should first remind ourselves of the fact that the symmetry properties in Eq. (\ref{symmetry}) can be observed in the model only under the condition of Eq. (\ref{assumption}). This symmetry is not present in the model of Eq. (\ref{Disc-Eneg-model-11}). To show the breakdown of the symmetry of Eq. (\ref{symmetry}) in the model of Eq. (\ref{Disc-Eneg-model-11}) in more detail, we replace the sum over triangles $\sum_{\it \Delta}$ of  $S_1$ and $S_2$ in Eq. (\ref{Disc-Eneg-model-11}) with the sum over bonds $\sum_{ij}$ before the condition of Eq. (\ref{assumption}) is imposed. In this new expression of $S_1$, which is expressed by the sum over  bonds $\sum_{ij}$, the Gaussian bond potential for the bond $12$, which is shared by the triangles ${\it \Delta}^+$ and ${\it \Delta}^-$ as in Figure \ref{fig-A2}(a) for example, is given by  
\begin{eqnarray}
\label{S1-on-12} 
S_1(\ell_{12})=(1/3)\left(\rho_{1}^++\frac{1}{\rho_{2}^+}+\rho_{2}^-+\frac{1}{\rho_{1}^-}\right) \ell_{12}^2\sim\gamma_{12}\ell_{12}^2.
\end{eqnarray}
In this expression, the former half 
$S_1^+(\ell_{12})\!=\!(1/3)\left(\rho_{1}^+\!+\!1/\rho_{2}^+\right) \ell_{12}^2$ is the contribution from ${\it \Delta}^+$, and the latter half 
$S_1^-(\ell_{12})\!=\!(1/3)\left(\rho_{2}^-\!+\!1/{\rho_{1}^-}\right) \ell_{12}^2$ is the contribution from ${\it \Delta}^-$. 
However, it is clear that $\gamma_{12}$ and hence $S_1(\ell_{12})$ in Eq. (\ref{S1-on-12}) are not symmetric under the change of surface orientation.  In fact,  we have 
\begin{eqnarray}
\label{S1-on-12-1} 
\bar S_1(\ell_{12})=(1/3)\left(\rho_{2}^++\frac{1}{\rho_{1}^+}+\rho_{1}^-+\frac{1}{\rho_{2}^-}\right) \ell_{12}^2\sim\gamma_{21}\ell_{12}^2
\end{eqnarray}
for the opposite orientation (see Figure \ref{fig-A2}(b)). In Eq. (\ref{S1-on-12-1}),  we write the coefficient of $\ell_{12}^2$ by $\gamma_{21}$ because it is obtained  from $\gamma_{12}$ in Eq. (\ref{S1-on-12}) by exchanging the suffixes $1$ and $2$. It is also easy to show that $\gamma_{12}\!\not=\!\gamma_{21}$ and hence  $S_1(\ell_{12})$ are not always identical to $\bar S_1(\ell_{12})$ in general.  

Thus, we find that the asymmetry $\gamma_{12}\!\not=\!\gamma_{21}$
means that $S_1(\ell_{12})$ is not invariant under the orientation
exchange. From this, we can  see that the Gaussian bond potential
energy (and also the bending energy) of the bond $12$ of one surface
configuration  differs from that of the opposite orientation
configuration.  However, we have no reason for the difference in $S_1$
for two surfaces with different orientations.  Thus, the model defined
by Eq. (\ref{Disc-Eneg-model-11}), which is orientation asymmetric, is
ill defined in the context of conventional surface modelling. 

Moreover, we have to remark that the model defined by
Eq. (\ref{Disc-Eneg-model-12}), which is orientation symmetric, is
also  ill defined. The reason for this ill definedness is that  the bond length squares calculated with $\rho^+$ in ${\it \Delta}^+$ is not always identical to the one calculated with $\rho^-$ in  ${\it \Delta}^-$ in Figure \ref{fig-A2}(a), where $\rho_1^\pm\!=\!\rho_2^\pm$ in the model of Eq. (\ref{Disc-Eneg-model-12}).  Indeed, the metric on  ${\it \Delta}^+$ is given by
$g_{ab}\!=\!\left(  
       \begin{array}{@{\,}ll}
        1/\rho^+ & \; 0 \\
        0 & \; \rho^+
       \end{array} 
       \\ 
 \right)$, where the coordinate origin is at the vertex $1$.  Then, we have the bond length squares  $ \left(1/\rho^+\right)\ell_{12}^2$ for the bond $12$ with respect to the metric $g_{ab}$, and changing the vertex origin from $1$ to $2$, we also have $ \rho^+\ell_{12}^2$. Thus,  summing over these two expressions without the coefficient $1/2$, we have  $\left(1/\rho^+\!+\!\rho^+\right)\ell_{12}^2$ for the bond length squares. Through the same procedure, we have  $\left(1/\rho^-\!+\!\rho^-\right)\ell_{12}^2$ from  ${\it \Delta}^-$. These two square lengths of the bond $12$ must be the same. However, we have
\begin{eqnarray}
\label{constraint} 
\frac{1}{\rho^+}+\rho^+\not=\frac{1}{\rho^-}+\rho^-
\end{eqnarray}
because $\rho^+\!\not=\!\rho^-$ in general. The edge length of
triangles should uniquely be given as the basic requirement even in
the discrete models. Therefore, in a model construction on the
triangulated lattices, we always obtain an ill-defined discrete model
if we start with an arbitrary Riemannian metric in which the elements
are defined on the triangles. Note that the bond "length" used here is
the length with respect to $g_{ab}$ on  ${\it \Delta}^\pm$ and is
different from the Euclidean bond length $\ell_{ij}$ (also note that $g_{ab}$ is simply a Riemannian metric at this stage).

However, these ill-defined models in Eqs. (\ref{Disc-Eneg-model-11})
and (\ref{Disc-Eneg-model-12}) become well defined in the context of Finsler geometry \cite{Koibuchi-Sekino-2014PhysicaA,Matsumoto-SKB1975,Bao-Chern-Shen-GTM200}.  In this context,  the bond length calculated with $g_{ab}$ on the triangle ${\it \Delta}^+$ can be considered as the direction-dependent length from $1$ to $2$, and the one calculated  with $g_{ab}$ on ${\it \Delta}^-$ can be considered as the length from $2$ to $1$ (Fig. \ref{fig-A2}(a)). Therefore, the inequality in Eq. (\ref{constraint}) is satisfied in general.  Moreover, the aforementioned quantities $S_1^+(\ell_{12})$ and  $S_1^-(\ell_{12})$ in Eqs. (\ref{S1-on-12}) and (\ref{S1-on-12-1}) are meaningful because these quantities are also considered as direction dependent in the Finsler geometry context (Fig. \ref{fig-A1}(c)).

%%%%%%%%%%%%%%%%%%%%%%%%%%%%%%%%%%%%%%%%%%
%\bibliographystyle{mdpi}

%=====================================
% References, variant A: internal bibliography
%=====================================
%\renewcommand\bibname{References}

%=====================================
% References, variant B: external bibliography
%=====================================
%\bibliography{your_external_BibTeX_file}

%%%%%%%%%%%%%%%%%%%%%%%%%%%%%%%%%%%%%%%%%%
%% optional
%\sampleavailability{Samples of the compounds ...... are available from the authors.}

%%%%%%%%%%%%%%%%%%%%%%%%%%%%%%%%%%%%%%%%%%
\end{document}